\documentclass[]{interact}

\usepackage{epstopdf}
\usepackage[caption=false]{subfig}
\usepackage{graphicx}%
\usepackage{multirow}%
\usepackage{amsmath,amssymb,amsfonts}%
\usepackage{amsthm}%
\usepackage{mathrsfs}%
\usepackage[title]{appendix}%
\usepackage[dvipsnames]{xcolor}%
\usepackage{textcomp}%
\usepackage{manyfoot}%
\usepackage{booktabs}%
\usepackage{algorithm}%
\usepackage{algorithmicx}%
\usepackage{algpseudocode}%
\usepackage{listings}%
\usepackage{gensymb}
\usepackage{lscape}
\usepackage{dsfont}
\usepackage{pdflscape}

\usepackage[numbers,sort&compress]{natbib}
\bibpunct[, ]{[}{]}{,}{n}{,}{,}
\renewcommand\bibfont{\fontsize{10}{12}\selectfont}

\theoremstyle{plain}

\theoremstyle{definition}

\theoremstyle{remark}

\def\R{{\mathbb R}}

\def\W{{\mathcal W}}
\def\V{{\mathcal V}}

\numberwithin{equation}{section}
 \DeclareMathOperator*{\argmax}{arg\,max}

\begin{document}

\articletype{ARTICLE TEMPLATE}

\title{Model selection for extremal dependence structures using deep learning: Application to environmental data.}

\author{
\name{Manaf  Ahmed \textsuperscript{a,b}\thanks{CONTACT Manaf ~Ahmed.  Email: manaf.ahmed@uomosul.edu.iq},  V\'eronique Maume-Deschamps\textsuperscript{b}  and Pierre Ribereau\textsuperscript{b}}
\affil{\textsuperscript{a}Department of Statistics and Informatics, University of Mosul, Al Majmoaa Street,41002, Mosul, Iraq; \textsuperscript{b}Universite Claude Bernard Lyon 1, CNRS, Ecole Centrale de Lyon, INSA Lyon, Université Jean Monnet, ICJ  UMR5208, 69622, Villeurbanne, France}
}

\maketitle

\begin{abstract}
This paper introduces a new methodology for extreme spatial dependence structure selection. It is based on deep learning techniques, specifically  Convolutional Neural Networks  - CNNs. Two schemes are considered: in the first scheme, the matching probability is evaluated through a single CNN while in the second scheme, a hierarchical procedure is proposed: a first CNN is used to select a max-stable model, then another network allows to select the most adapted covariance function, according to the selected max-stable model. This model selection approach demonstrates performs very well on simulations. In contrast, the Composite Likelihood Information Criterion CLIC faces issues in selecting the correct model.  Both schemes are applied to a dataset of 2m air temperature over Iraq land,  CNNs are trained on dependence structures summarized  by the Concurrence probability.
\end{abstract}


\section{Introduction}\label{sec1}
Model selection is an essential task in statistical modelling to provide  reliable, interpretable, and predictable models. We are especially interested in  environmental and climatic  phenomena. Statistical approaches for model selection often rely on full likelihood framework and use information criteria such as  Akaike, Takeuchi, and Bayesian  Information Criteria (reps. AIC, TIC, BIC).\\ 
In spatial contexts, some investigations on model selection criteria have been conducted, for instance, in \cite{hoeting2006model} an information criterion constructed on a heuristic derivation of AIC  to be applicable for spatial models, named the corrected Akaike Information Criterion AICc is proposed. A general methodology is proposed in \cite{Huang2007} for model selection, it is based on an unbiased estimator of mean square prediction errors. The performances of BIC, AIC, and AICc in spatial context assessed in \cite{HyeyoungPerformance2009} and \cite{wahyu2024applying} reveal some lack of robustness for AIC and BIC, depending on stationarity, isotropy and sample size. We are concerned with spatial extremes whose statistical inference is still challenging, see \cite{Blanchet2018}. In the context of spatial max-stable processes, the full likelihood is not computable. The composite likelihood estimation method and  the Composite Likelihood Information Criterion CLIC for model selection are adapted (see \cite{padoan2010likelihood} and  \cite{Cristiano2005}). In \cite{Thibaud2013}  CLIC is shown  to be close to AIC, a tuneable model selection criterion is introduced in \cite{castilla2020model}, driven from the density power of  classical dissimilarity measures.  Also, a simulation-based approach using approximate Bayesian computation was performed as a model-selection for max-stable spatial processes in \cite{LEE2018128}.\\

\begin{figure}[htb]
 \centering
\includegraphics[width=1\textwidth]{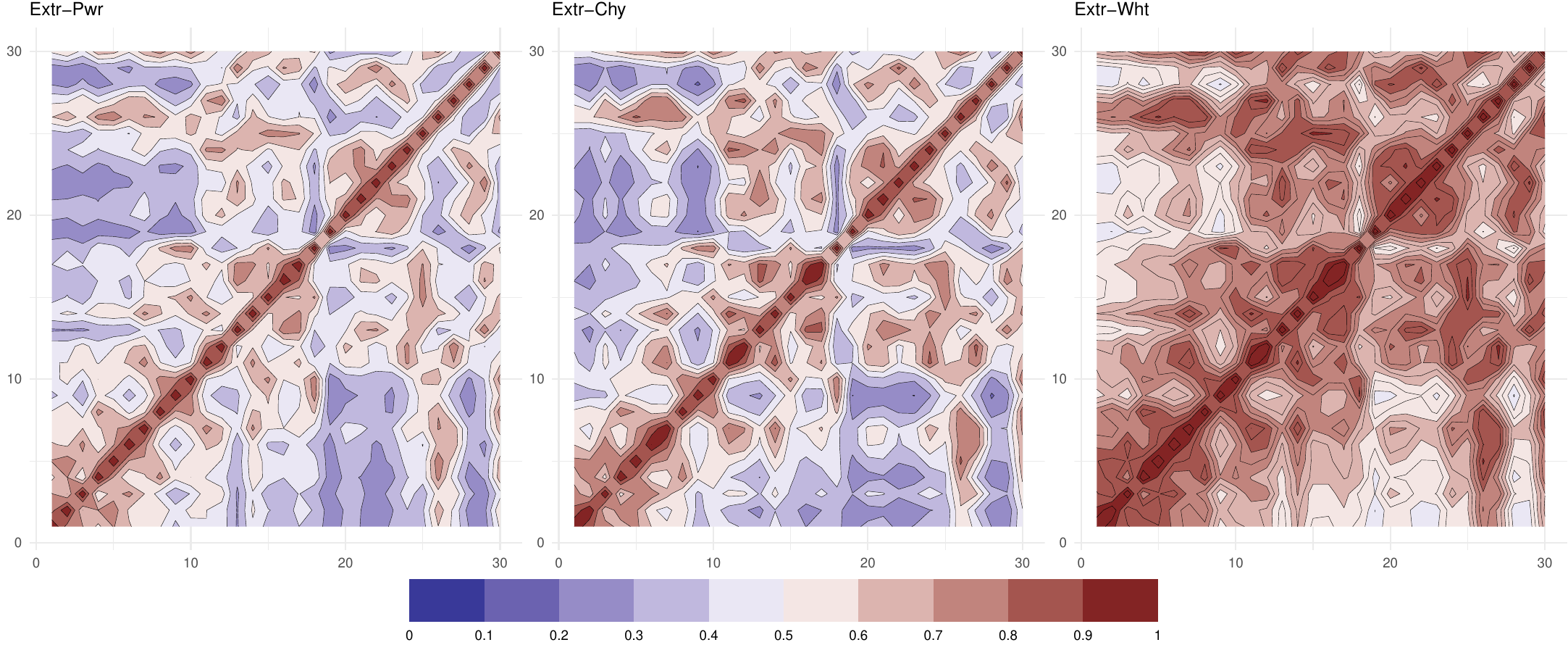}
  \caption{Pairwise concurrence probabilities $Con(s_i,s_j), i,j=1,\cdots,30$ for different Extremal-t max-stable processes. The panels in the figure from left to right are: Extremal-t model with Power covariance function, Extr-Pwr;  Extremal-t model with Cauchy covariance function, Extr-Chy;  and Extremal-t model with Whittle-Mart\'ern covariance function, Extr-Wht. The models are generated randomly on $s_i, i=1,\cdots,30$ fixed locations in $[0,1]^2$. In addition,  the parameters are fixed: $\sigma=1.25$, $k=1.5$ and $v=3$  for the three models.\label{fig:1}}
\end{figure}

 Despite the classical information criteria, especially CLIC  extensively used by statisticians in the model selection of spatial extremes, the complexities in the statistical inference of these models are obstacles to the performance of these criteria, e.g, a weak efficiency in selecting the correct model in a simulation study is recorded  in\cite{padoan2010likelihood}. Moreover, in \cite{davison2013geostatistics}  it is observed that this criterion may not be decisive enough to select the correct model. These fluctuations in the performance of classical model selection criteria encouraged us to consider another methodology. Usually, most environmental phenomena are characterised by their spatial features. Hence, if the spatial features of a theoretical model match the spatial features of the phenomenon more than others, this model will be more representative of the phenomenon. This point of view  was the motivation behind proposing a deep learning approach based on Convolutional Neural Networks -   CNN - going one step further than \cite{https://doi.org/10.1002/env.2714}. In \cite{https://doi.org/10.1002/env.2714}, CNNs were used to determine whether an extremal dependence structure is asymptotically independent or asymptotically dependent or mixed. Now, we aim to propose a methodology for selecting one max-stable process. There are variations in shapes and levels for the strength of dependence of spatial dependence structures among the different max-stable models, e.g, in Figure~\ref{fig:1}.  A spatial dependence measure able  to capture spatial characteristics of the dataset will be used in order to investigate  representative models, rather than using the raw data directly. This allow to achieve higher performance of neural networks and to shorten the training time for these networks. The concurrence probability dependence measure is more compatible with our objective.\\

The paper is organized as follows. In Section \ref{sec:2}, we introduce  statistical tools for spatial max-stable processes used in this study, including  the spectral representation of such models, the concurrence probabilities, and the composite likelihood estimation method. Section \ref{sec:3} is devoted to the statistical framework of the convolutional neural network. An overview of the 2m air temperature over the Iraq land dataset, fitting models to this data, visualization of these fitted models, and building the corresponding CNNs are presented in Section \ref{sec:4}. Section \ref{sec:5} is devoted to a complete presentation of the methodology that we have used in the previous section. Section \ref{sec:6} presents some for discussions and conclusions. Finally,  some additional results are presented in the Appendix. 

\section{Statistical  tools of max-stable spatial process  }\label{sec:2}
We provide a brief overview of the definitions and properties of spatial max-stable processes and related statistical tools. 
\subsection{Spatial max-stable  processes}
Spatial max-stable processes are widely used for studying  environmental extremes. The well-known spectral representation introduced in \cite{laurens2006} is crucial to construct spatial extremes.  Let $\{X'_k(s)\}_{s\in\mathcal{S}}, \mathcal{S}\subset\mathbb{R}^d$, $d\geq1, k=1,2,\cdots$ be  i.i.d copies of a stochastic process  with continuous sample paths. If there exist  sequences $a_n(s)>0$ and $b_n(s)$, $n\in\mathbb{N}$, such that  
\begin{equation}
\bigg\{\max_{k=1,\cdots,n}\bigg(\frac{X'_k(s)-b_n(s)}{a_n(s)}\bigg)\bigg\}_{s\in\mathcal{S}}\overset{d}{\to}\{X(s)\}_{s\in\mathcal{S}}
\end{equation}
as $n\to\infty$ with non-degenerate limiting distribution, then $X:=\{X(s)\}_{s\in\mathcal{S}}$ is a spatial max-stable process whose marginal laws are Generalized Extreme Value distributions  GEV, i.e, for any ${s\in\mathcal{S}}$, 
\begin{equation}
 G_{X(s)}(x)= \exp\bigg\{-\bigg[1+\xi(s)\bigg(\frac{x-\mu(s)}{\sigma(s)}\bigg)\bigg]^{-1/\xi(s)}\bigg\},
\end{equation}
where $\mu$, $\sigma$, and $\xi$ respectively are  location, scale, and shape parameters (see \cite{dehaan1984}).  The process $\{X(s)\}_{s\in\mathcal{S}}$ is  a simple  spatial max-stable process if its margins are unite Fr\'echet, i.e, $G_{X(s)}(x)=\exp\{-1/x\}$, $ \forall s\in\mathcal{S}$,   the parameters  $\mu(s)=\sigma(s)=\xi(s)=1$.  Any simple spatial max-stable process  $\{X(s)\}_{s\in\mathcal{S}}$ admits a spectral representation as presented in  \cite{laurens2006}, in the sense that it exists $\{\eta_i,i\geq1\}$, a i.i.d point Poisson process on $[0,\infty)$, with intensity $d\eta/\eta^2$, and  $\{Z^+_i(s)\}_{s\in\mathcal{S}}$  i.i.d copies of  the non-negative part of a  stochastic process $Z$, with  $\mathbb{E}[Z^+(s)]=1$, for all $s\in\mathcal{S}$, and $Z^+(s)=\max\{Z(s),0\}$, such that
\begin{equation}\label{spect:rep}
X(s)=\max_{i\geq 1}\eta_iZ^+_i(s), 
\end{equation}
The multivariate  distribution function  writes
\begin{equation}\label{expo:mesu}
\mathbb{P}\big(X(s_1),\cdots,X(s_d)\big)=\exp\big\{-V(x_1,\cdots,s_d)\big\},
\end{equation}
where $V$ is  called the exponent measure and writes 
$$V(x_1,\cdots,x_d)=\mathbb{E}\bigg[\max_{i=1,\cdots,d}\frac{Z(s_i)}{x_i}\bigg].$$ 
The exponent measure $V$ satisfies the  inequalities
$$ \max_{i=1,\cdots, d} \{1/x_i\} \leq V(1/x_1,\cdots,1/x_d)\leq \sum_{i=1,\cdots, d}1/x_i,$$
where the lower and upper bounds  are respectively reached for complete dependence resp. independence \cite{wadsworth2012dependence}. 
The flexibility of the spectral representation in  (\ref{spect:rep}) arises from the ability to consider  different models for the stochastic process $Z(s)$.   Smith model   (\cite{smith1990max}) was the first model introduced by considering  $Z_i(s)=\phi(s-s_i;\Sigma)$, where $\phi$ is the p.d.f of a $d$ dimensional Gaussian process with mean equal to zero and covariance matrix $\Sigma$. If $diag(\Sigma)=\sigma^2$, then $X(s)$ is an isotropic process. Brown-Resnick model is widely used, it was  firstly introduced in \cite{brown1977extreme} and its spectral representation is described in \cite{kabluchko2009stationary}: $X(s)=\max_{i\geq 1}\eta_i\exp\{\epsilon(s)-\gamma(s)\}$, where $\epsilon(s)$ is  a stationary Gaussian process with $\mathbb{E}[\epsilon(s)]=1$, and Variogram $\gamma(h)$, where $(s,s^*)\in\mathcal{S}\subset\mathbb{R}^2$, and $h=||s-s^*||_2$. Schlather model is constructed in \cite{schlather2002models} via  the  spectral representation: $X(s)=\max_{i\geq 1}\eta_i\sqrt{2\pi}\max\{0,\epsilon_i(s)\}$, where $\epsilon$ is standard Gaussian process with correlation function $\rho(h)$. For all the previous models $Z(s)$ is a  Gaussian process, while extremal-t processes (\cite{opitz2013extremal})  are constructed from a student distribution:   $X(s)=\max_{i\geq 1}\eta_i\beta_v\max\{0,\epsilon(s)\}^v$, with   $$\beta_v=\sqrt{\frac{\pi}{2^{(v-2)}}}\Gamma{\bigg(\frac{v+1}{2}\bigg)}^{-1}, v\geq 1,$$ $\epsilon$ as for Schlather models,  $v$ is the degree of freedom and $\Gamma$ the a gamma function. 

If the variogram of a Brown–Resnick process satisfies: $\gamma(h)\propto h^2$, $h\geq 0$, then it coincides with  the isotropic Smith model (\cite{davison2012statistical}). Moreover,  an Extremal-t  model tends to  a Schlather  model when  $v\to1$, while  the same model tends to a Brown–Resnick model as $v\to\infty$ (\cite{opitz2013extremal}).  Also, a  Whittle-Mat\'ern covariance function $\rho_{wht}(h;\lambda,k)$ tends to an exponential covariance function $\rho_{exp}(h;\lambda)$ when  $k\to 0.5$;  a power exponential covariance function $\rho_{pwr}(h;\lambda,k)$ tends to an exponential covariance function $\rho_{exp}(h;\lambda)$ (respectively, Gaussian covariance function $\rho_{gau}(h;\lambda)$) when $k\to1$(respectively, $k\to2$) (\cite{abrahamsen1997review}). These relations between different max-stables models may make the identification difficult.

\subsection{Concurrence probability dependence measure }
Our aim is to use CNN for model selection. Using the raw data as inputs for the CNNs may not serve the goal of this paper for two reasons: the confusion of CNN in extracting the spatial features and the time consumption in the training process. Leveraging the advantage of the power of CCNs requires highlighting the spatial patterns and extracting these patterns from the data. Consequently, investigating which tools are efficient and appropriate to extract these features is fundamental to minimising the loss of CNN. In addition, the number of tools that will be used also affects the performance of CNN loss. For example, in  \cite{https://doi.org/10.1002/env.2714}, CNN outperformed when using two dependence measures to construct the dependence structure of the datasets for identifying asymptotic dependence and independence models. Our current purpose is different and requires a dependence measure sensitive to slight differences occurring in extremes. The ability of the concurrence probability dependence measure to extract  the spatial patterns from the raw datasets motivated  the use of  this measure in summarizing the raw datasets. The power of this tool makes it sufficient without any complementary tools. On another hand, the CNNs will be trained on datasets with only one tensor.\\
To define the concurrence probability, let $\phi_i=\max_{\phi\in\Phi}\xi_iZ_i^+(s)$, $\Phi=\{\phi_i,i\geq1\}$ be a rewritten  formula of  the  spectral representation of $\{X(s)\}_{s\in\mathcal{S}}$ in Equation \ref{spect:rep}. The extremes are  concurrent at $ s\in\mathcal{S}$, if  for the some i.i.d  copies $X_k(s)$ have the same spectral function $\phi$, i.e, for some $k\geq1$, we have 
\begin{equation}
X(s)=\phi_k(s), \quad   s\in\mathcal{S}
\end{equation}
with concurrence probability of extremes
$Con(s_1\cdots,s_n)=\mathbb{P}r\big(\text{for some} \quad k\geq1, X(s)=\phi_k(s)\big).$
According the Theorems 2  in \cite{doi:10.1080/01621459.2017.1356318}, $Con(s_1\cdots,s_n)$  can written as   
\begin{equation}\label{con:max}
Con(s_1,\cdots,s_k)=\sum_r^k (-1)^r\sum_{j\subseteq \{1,\cdots,k\}; |J|=r} \mathbb{E}\bigg[\log\mathbb{P}\{X(s_j)\leq x(s_j), j\in J\}\bigg], 
\end{equation} 
where $x(s)$ is the independent copy of $X(s)$. For any  pairwise  $(X(s_i),X(s_j))$, $(s_i,s_j)\in\mathcal{S}$ case, by Theorems 3  in \cite{doi:10.1080/01621459.2017.1356318}, the extremal concurrence probability equals  the Kendall correlation coefficient, such that  
\begin{equation}\label{kandall}
\begin{split}
Con(s_i,s_j)=&2+ \mathbb{E}\bigg[\log\mathbb{P}\{X(s_i)\leq x(s_i),X(s_j)\leq x(s_j)\}\bigg]\\
= & \{ \text{sign}(X(s_i)-x(s_i))\text{sign}(X(s_j)-x(s_j))\}\quad \in[0,1]. 
\end{split}
\end{equation} 
Note that $X(s_i),X(s_j)$ are completely independent (resp. dependent), if $Con(s_i,s_j)=0$ (resp. $Con(s_i,s_j)=1$ ). Equation \ref{kandall} is used to construct the pairwise dependence structure of $X$.
\subsection{Inference and  model selection of max-stable spatial processes}\label{sec3}
 Inference of spatial max-stable processes using the likelihood is not accessible because of computational complexity. In \cite{padoan2010likelihood} it is proposed to use the composite likelihood. The dimension $2$ composite likelihood is the most used for inference of max-stable models. It writes: 
$$\mathcal{L}(\psi)=\sum_{\forall k\in K}\sum_{\forall i\in \mathcal{S}}\sum_{\forall (j>i)}\log f(x^k_i,x^k_j;\psi),\quad (i,j)\in\mathcal{S}, \mathcal{S}\subset\mathbb{R}^2, \text{and} \quad k=1,\cdots,K,$$
where $\psi \in \Theta$ is the vector of parameters. The corresponding Composite Likelihood Information Criterion CLIC is given  by 
\begin{equation}\label{clic}
CLIC=-2\bigg[\mathcal{L}(\hat{\psi})-tr(J^{-1}(\hat{\psi})K(\hat{\psi}))\bigg],
\end{equation}
 where $\hat{\psi}_F=\arg\max_{\psi\in\Theta}$, $J(\psi)=\mathbb{E}[-\nabla^2\mathcal{L}(\psi)]$, and $K(\psi)=Var(\nabla\mathcal{L}(\psi))$ are receptively the sensitivity  and variability matrices. This selection criterion used also with weighted composite Likelihood estimation introduced in \cite{Gaetan2012}. 
  
\section{ Statistical framework of the Convolutional Neural Networks CNNs }\label{sec:3}
The Convolutional Neural Network CNN model is one of the significant artificial intelligence techniques to construct the features of patterns from neighbours in many topics, such as image processes, Geographic Information System GIS, and environmental applications. A review of the statistical framework of neural networks can be found in \cite{wikle2023statistical}. Recently, the availability of spatial or spatiotemporal datasets and the ability of CNN techniques to extract spatial patterns encouraged the statistical community to employ these techniques to address questions which are still hardly treated by classical statistical tools. For example, in \cite{https://doi.org/10.1002/env.2714} a CNN was designed to discriminate between asymptotic dependence and asymptotic independence  by considering two statistical dependence measures.   A hybrid estimation method was proposed in \cite{walchessen2024neural}  by training CNNs  on the likelihood function to model the extremes.\\

Mainly the CNN model $\mathcal{G}:\V_1\times (\W_1\times \cdots \times \W_L)\to \V_{L+1}$ is a composition of $i\in [L], L\in\R^+$ layerwise functions $g_i:\V_i\times\W_i\to\V_{i+1}$:
\begin{equation}
\mathcal{G}(z;K)=\big(g_L\circ \cdots \circ g_1\big)(z),
\end{equation} 
where $z_i\in\V_i$ and $K_i\in\W_i$, respectively are the state and parameters variables such that  $\{(K_i,K_j):  i\neq j \}$ are independent. In a CNN,  a layer $i\in [L]$ may be one of  the three essential types,  convolutional, fully connected, and polling. In order to reduce the dimension of $z$ produced by the convolutional layers, the pooling layers are used. The convolutional and fully connected layers contain trainable parameters while pooling layers have no parameters. 
In the convolutional layers, the  features are extracted by a kernel $K$:  for each layer $i\in[L]$, the layerwise function mapping  is given by   
 \begin{equation}\label{layer-func}
g_i:\big(\R^{n_i\times m_i}\otimes\R^{\ell_i}\big)\times \big(R^{p_i\times q_i}\otimes\R^{\ell_{i+1}}\big)\to \big(\R^{n_{i+1}\times m_{i+1}}\otimes\R^{\ell_{i+1}}\big),
\end{equation}   
and the  output of convolutional block action is
 \begin{equation}
z_{i+1}:=f_i(z_i,K_i)=\Psi_i\big(\Phi_i\big(\mathcal{C}_i(z_i,K_i)\big)\big)\in\R^{n_{i+1}\times m_{i+1}}\otimes\R^{\ell_{i+1}},
\end{equation}
where $z_i\in\V_i, \V_i\in\R^{n_i\times m_i}\otimes\R^{\ell_i}$ is the  input of the convolutional layer $i$ with matrix size $(n_i,m_i)$.  The kernel  $K_i\in\W_i, \W_i\in\R^{p_i\times q_i}\otimes\R^{\ell_{i+1}}$ is a set of parameters which will be the tuned by the training process.\\

The convolutional operation between $z_i$ and $K_i$ in  $g_i$ will be done by the operator $\mathcal{C}_i$. The kernel $K_i$  moves across the feature map positions of $z_i$ with stride size  $\Delta\in\R^+$.  This operation aims to extract the features of the dependencies from the neighbour locations to several feature maps (filters). The number of neighbours is determined by the kernel size. The hyper-parameter $\Delta$, the size of kernel $(p_i,q_i)$,  and the depth of the feature map $\ell_{i+1}$ should be previously fixed. 
The elementwise function (activation function) $\Phi_i: \R^n\to\R^n$ will take the decision which the features maps extracted by $\mathcal{C}_i$ will be the output of the layer $i$. 
Mostly, the convolutional operation increases the dimensionality of the layer output and  increases extremely  the number of trainable parameters, which makes CNNs quite complicated to calibrate. For that, for every disjoint region of size $r\times r$ in each feature map in the some or all layers, the pooling operator  $\Psi_i:\R^{\hat{n}_{i}\times\hat{m}_{i}}\otimes\R^{\ell_{i+1}}\to\R^{n_{i+1}\times m_{i+1}}\otimes\R^{\ell_{i+1}}$ will reduce these dimensions to avoid the increase of the number of parameters, where $\R^{\hat{n}_{i}\times\hat{m}_{i}}$ denote to intermediate  inner space.  \\
The main mission of the conventional layers is to extract feature maps, not classify the categorical data. While the fully connected (dense) layer will classify the features. For that, the dense layers will be in the architecture of the CNN model. Let   $z_i^*\in \R^{n_i}$ be the feature map  obtained after converting the output of the previous 2D convolutional layer into a one-dimensional vector, commonly referred to as the flatten operation. Let  $W_i\in\R^{n_{i+1}\times n_i}$, and  $b_i\in\R^{n_{i+1}}$ be a weight matrix and bias vector, respectively. Then  the layerwise function $f_i:\R^{n_i}\times\big(\R^{n_{i+1}\times n_i}\times\R^{n_{i+1}}\big)\to\R^{n_{i+1}}$ for the dense layer $i\in L$ is 
\begin{equation}
f_i(z_i^*,W_i,b_i)=\Phi(W_i\cdot z_i^*+b_i). 
\end{equation} 
Generally, in the multi-categorial classification tasks, the activation function is chosen  as a softmax function:
 \begin{equation}\label{soft:max}
 \Phi(z^*_j)=\frac{\exp(z^*_j)}{\sum_{\forall j}{\exp(z^*_j)}}.
 \end{equation}    
The weights  $K_i$,  $W_i$, and $b_i$ are adjusted in the training phase, it requires a suitable loss function. The Kullback–Leibler divergence is one of the statistical distances that can be assumed to be used as a loss function. Let $q_{\psi}=F(x,\psi)$,  the Kullback–Leibler divergence loss function is given by 
\begin{equation}\label{KL}
\mathcal{KL}=-\sum_{\forall a\in \Omega}P(a)\log\bigg(\frac{P(a)}{Q_{\psi}(a)}\bigg),
\end{equation}
where $P$, and $Q$ respectively are the  true and predicted distributions of the class $a\in \Omega$. The training phase will minimize  $\mathcal{KL}$ by updating the weights  set $\{K_i, W_i, b_i\}$. More details concerning the mathematical framework of neural networks can be found in \cite{caterini2018deep}. The accuracy of the prediction model is the number of times the predicted model matches the true one, i.e, $ \mathcal{AC}=\#\{\hat{y}_k \equiv y\}/{N} $ with $\hat{y}:=\max_{a_i\in \Omega}\{Q(a_i)\}$.

\section{Two meter  air temperature over Iraq dataset}\label{sec:4}
\subsection{Data overview}
\begin{figure}[ht!]
  \centering
  \includegraphics[scale=0.52]{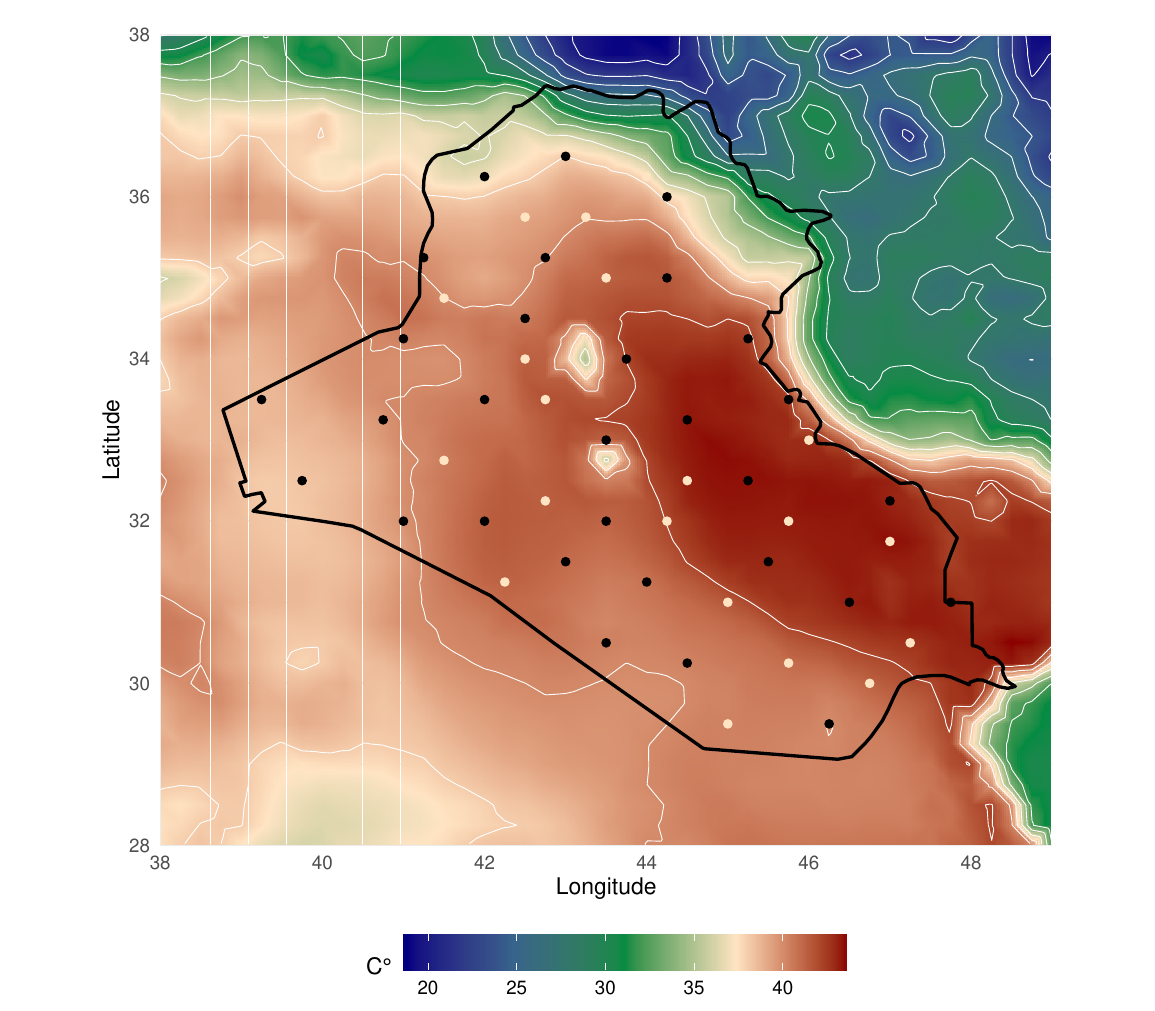}
  \caption{presents the average of the 2m air temperature in the period from 1980 to 2023 for June, July, and August and during the hours between 11:00 and 17:00 inside the coordinates longitude  $(38 -49){\degree}$ and latitude $(28-38){\degree}$ with grid size $11$ km$^2$. The dataset in the locations with the black points (30) is used in the modelling and training, while the dataset in the locations with white points (20) is devoted to verifying the matching of the selected estimated model with the first one. \label{fig:2}}
\end{figure}

We shall now present a selection model based on CNN for two meters air temperature in Iraq  $\{Y(s_i)\}_{s\in\mathcal{S}}$, $\mathcal{S}\subset\mathbb{R}^2$ from 1980 to 2023 in the summer period (June, July, and August)  during the hours between 11:00 and 17:00 were selected from $|\mathcal{S}|=45\times 41$ grids with size  $11$km$^2$. This grid is between  longitude $(38 -49){\degree}$ and latitude $(28-38){\degree}$ so that covers Iraq land. The dataset is provided by the meteorological reanalysis  of the European Center for Medium-RangeWeather Forecasts ECMWF, called the ERA5 dataset.  Fifty grids $s_i,i=1,\cdots,50$, $s\subset\mathcal{S}$, (30 for modelling and 20 for validation) were chosen randomly as shown in Figure \ref{fig:2}.  
The mountains in the north of Iraq are excluded from the modelling in order to ensure spatial stationarity. The same dataset has been used in \cite{https://doi.org/10.1002/env.2714} with some modifications in the years.  The same investigation is considered here: the examination of the temporal stationarity and the isotropy property. The moving average smoothness of the spatial gradient was applied over $10$ days before dealing with the dataset. That is $\forall s\in\mathcal{S}$, $\hat{Y}_k(s)=Y_k(x)-\hat{\mu}_k(s)$  is the smoothed stochastic  dataset with moving average $\hat{\mu}(s)$ over $10$ days.  Monthly block maxima are considered for each of the selected points of the grid, so that $$\{X(x)\}_{s\in\mathcal{S}}=\max_{k \in N_m(k)}\{\hat{Y}_k(s)\},\quad N_m(k)=\{k^*:k-m<k^*< k+m\},$$ where $N_m(k)$ is a non-overlapping set of temporal neighbours of $k$ with size $m$. For more information, see  \cite{https://doi.org/10.1002/env.2714} and \cite{huser2021eva}.The size of neighbours in this application has been chosen as $m=31$ for the months of June and August, and $m=30$ for July results $k=132$ replicate. 
\subsection{Modelling the dataset}\label{modeling}
Modelling 2m air temperature requires transferring  monthly block maxima  $\{X(s)\}$ to  max-stable process with unit Fr\'echt margin for each site in the dataset. We have used empirical transformation  i.e. $\hat{X}(s_i)=-1/\log{\tilde{G}(X(s_i))}$. We propose  17 combinations of max-stable models as candidates in order to select the best model among them. Proposing max-stable models is in accordance by the results in  \cite{https://doi.org/10.1002/env.2714}. The dataset shows  asymptotic dependence  behaviour. Models are  Smith, and Brown-Resnick with variogram $\gamma(h)=||{h}/{\lambda}||^k$, and components of max-stable models from Schlather, Geometric, and Extremal-t  with five covariance functions.  The covariance functions proposed are Exponential, $\rho_{exp}(h)=\exp\{{-h}/{\lambda}\}$; Gaussian, $\rho_{gau}(h)=\exp\{- (h/{\lambda})^2\}$; Power exponential,  $\rho_{pwr}(h)=\exp\{- ({h}{/\lambda})^k\}$; Cauchy, $\rho_{chy}(h)=\{1+({h}/{\lambda})\}^{-k}$; and  Whittle-Matern  $\rho_{wht}(h)= \frac{2^{1-k}}{\Gamma{k}}\big(\frac{h}{\lambda}\big)^k \mathcal{K}_k\big(\frac{h}{\lambda}\big) $, where $\lambda$ and $k$ respectively are the scale and smooth parameters.

\begin{table}[htb]
\caption{This table presents the parameter's estimation, likelihood amounts $\mathcal{L}$, CLIC, and model ranks. The notations BR refers to Brown-Resnick; Schl-Exp to  Schather with  $\rho_{exp}$; Schl-Gau to Schather with $\rho_{gau}$;  Schl-Pwr to Schather with  $\rho_{pwr}$; Schl-Chy to Schather with  $\rho_{chu}$; Schl-Wht Schather with  $\rho_{wht}$; Extr-Exp to  Extremal-t with  $\rho_{exp}$; Extr-Gau to Extremal-t with $\rho_{gau}$;  Extr-Pwr to Extremal-t with  $\rho_{pwr}$; Extr-Chy to Extremal-t with  $\rho_{chu}$; Extr-Wht Extremal-t with  $\rho_{wht}$; Geom-Exp to  Geometric with  $\rho_{exp}$; Geom-Gau to Geometric with $\rho_{gau}$;  Geom-Pwr to Geometric with  $\rho_{pwr}$; Geom-Chy to Geometric with  $\rho_{chu}$; Geom-Wht Geometric with  $\rho_{wht}$. 
}\label{tab:1}
\small
\begin{tabular}{@{\extracolsep{-5pt}}lcccccccc}
 \toprule
   \multirow{2}{*}{Model }   &   &    \multirow{2}{*}{Scale $\hat{\lambda}$}  & \multirow{2}{*}{Smooth $\hat{k}$} &   \multirow{2}{*}{Degree of freedom $\hat{v}/ \hat{\sigma}$} & & \multirow{2}{*}{Likelihood $\mathcal{L}$} &   \multirow{2}{*}{CLIC}  & \multirow{2}{*}{ ${Rank}$ }  \\ 
                                                                                                                                                                                           &  &  & &&  &    \\
   \midrule  
Smith&                                    & 0.3488                      & --------                   &--------           && 199888.8            & $399904.2$   &      17 \\
BR &                                       & 1.4718                   & 1.2208             & --------                  && 197992.3             & $396242.6$    &     11 \\
   \midrule  
Schl-Exp  &                            & 3.8379                 &--------                 &--------                    && 195123.4            & $390515.0$      &   9 \\
Schl-Gau &                             & 1.2192                   & --------                 &--------                &&195645.2             & $391537.5 $        &10  \\
Schl-Pwr &                             & 2.2596                 & 1.3053              & --------                   &&194996.1  &          $ 390279.5 $         & 7 \\
Schl-Chy &                             & 0.2878            & 0.1056              & --------                        && 195032.0             & $390350.5$       &  8  \\
Schl-Wht &                             &1.8796                   & 0.7217 & --------                               && 194995.8 &          $390277.5$  &6 \\
   \midrule  
Geom-Exp &&  97.594    & --------     &49.982                                                                    && 198139.6          & 396546.4 & 15\\
Geom-Gau&&  0.4382    &---------     & 0.4267                                                                   && 198334.6           & 396922.6   & 16\\ 
Geom-Pwr &&4.1631 &1.2532   &  3.9639                                                                        && 197993.9            &396259.1& 13\\
Geom-Chy && 0.2576&  0.0445&4.2084                                                                           && 198102.8& 396476.2&14 \\
Geom-Wh && 14.964  &0.6219 &17.360                                                                           &&197993.0&396257.1 &12 \\
   \midrule  
Extr-Exp &                              & 8.0366                  & --------                  &2.2513               &&194061.9             & $388498.6$      &4  \\
Extr-Gau &                              & 1.7565                   &--------                & 2.2214               && 194567.6            & $389448.1 $       & 5 \\
Extr-Pwr &                               & 3.7931              & 1.3453                 &2.3169                 &&193868.4            & $388124.2 $       &   2\\
Extr-Chy &                               & 0.3112           & 0.0545    & 2.2976                                 && 193920.1          & $388225.7 $ &3 \\
Extr-Wht &                               & 3.3068  & 0.7281 & 2.3154                                             &&193868.8 &       $\textbf{388124.0}$  &1\\ 
 \bottomrule
\end{tabular}
\end{table}
The best-ranked models are the component models of the Extremal-t. The flexibility of this model leads CLIC to choose it. Regarding the Extremal-t model, Extr-Wht is the best-fitted model with a slight difference in the CLIC with the Extr-Pwr. As well as in Schalther, Schl-Wht and Schl-Pwr, and Geometric Geom-Wht and Geom-Pwr. Recall that $\rho_{wht}\to \rho_{exp}$ as $k\to 0.5$ so that the slight differences in estimated parameters and the CLIC indicate that these models are close. For this reason,  Schl-Pwr, Geom-Pwr, and Extr-Pwr, have been removed (since they have the worst CLIC) in the next step of selecting the best representative model.

\subsection{Visualization the fitted models}\label{sec:vis}
\begin{figure}[htb]
  \centering
 \includegraphics[width=1\textwidth]{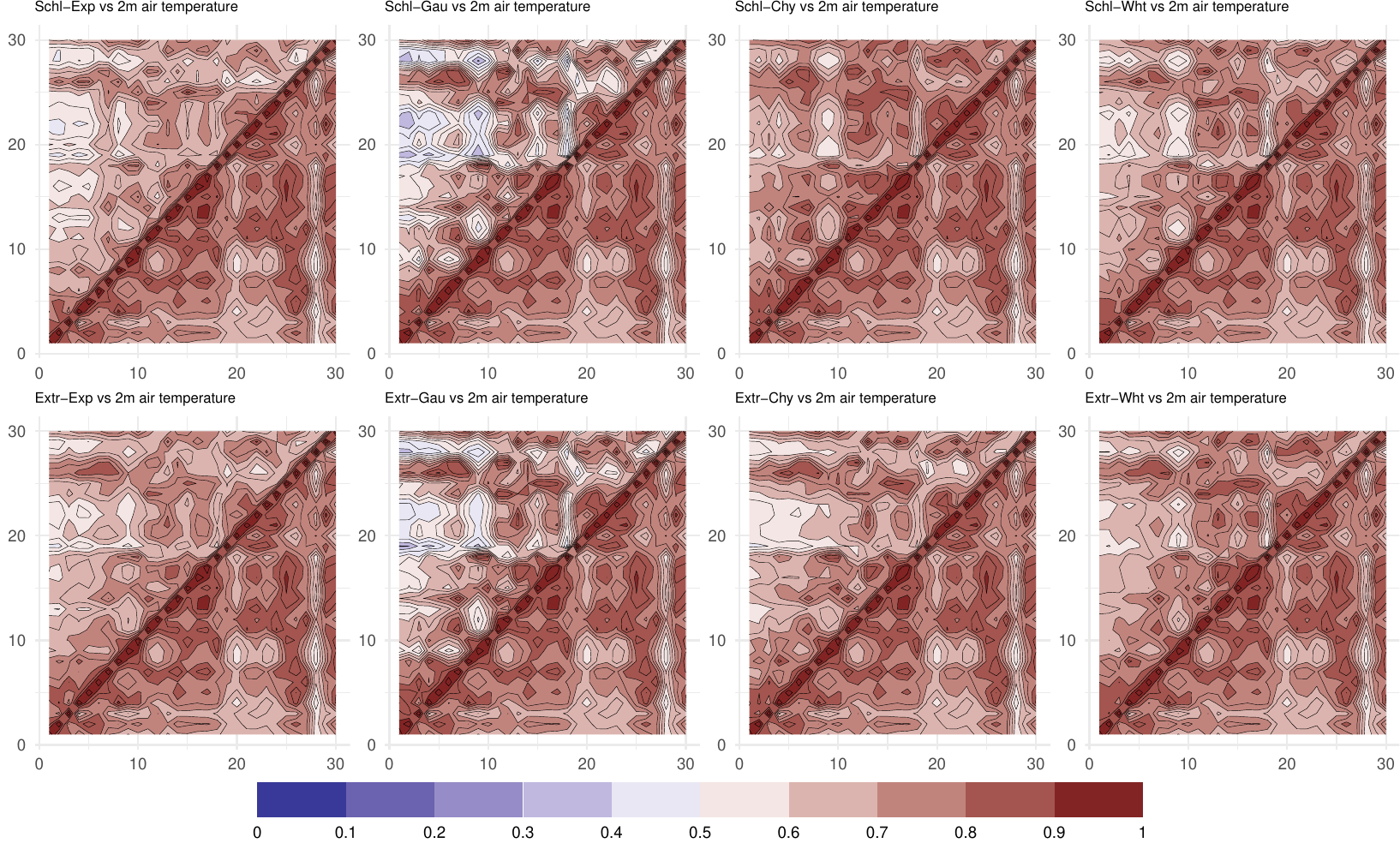}
  \caption{presents the dependence structures of the 2m air temperature dataset with the datasets simulated by the best eight fitted models (Schl-Exp, Schl-Gau, Schl-Chy, Schl-Wht, Extr-Exp, Extr-Gau, Extr-Chy, and Extr-Wht) with the same coordinates of 30 locations and number of the block maxima. These dependence structures are constructed by the pairwise  Concurrence probability dependence measure ${Con}(x_i,x_j)$, $i,j=1,\cdots,30$  in Equation \ref{kandall}. The upper triangle represents the pairwise $Con$ of the fitted model, while the lower triangle consists of the pairwise $Con$ of the 2m air temperature dataset.\label{fig:3}}
\end{figure}

A way to visualize the matching quality is to compare the estimated pairwise concurrence probability on the data and the estimated concurrence probability on simulated data for the fitted models. In each penal in Figure  \ref{fig:3}, the upper triangle consists of the estimated concurrence probability of the simulated models, while the lower triangle for all the panels represents the 2m air temperature. 

Firstly, the strength of the dependence is strong. This may increase the difficulty in selecting a best-fitted model. Generally, despite recognizing the symmetry in patterns of the 2m air temperature dataset with the models is not easy, it seems that the models Schl-Exp, Schl-Gau, Extr-Exp, and Extr-Gau are far from the observed dataset. Meanwhile, the models Schl-Chy, Schl-Wht, Extr-Chy, and Extr-Wht, especially the latter model (first ranked by CLIC) are closer to the data but is  quite hard to recognize which one is more symmetric with the 2m air temperature dataset. We shall discriminate between these models using CNN.

\subsection{Building and training CNNs }\label{sub:4}
Two schemes are considered for selecting the models. The first scheme is a one-step selection model by using  a single CNN.  In other words, a single CNN will be  trained on the set of the 14 models, say $\Omega_s$, used in Section \ref{modeling} excluding the three models Schl-Pwr, Geom-Pwr, and Extr-Pwr as mentioned previously. This trained CNN will provide the probability strength that the selected model matches the 2m air temperature dataset, i.e, $\Phi_s(z)\in[0,1], z\in\Omega_s$, $\Omega_s=\{\text{Sm}, \text{BR},\cdots, \text{Geom-Wht}\}$, where $\Phi\in[0,1]$ is the softmax activation function defined in \ref{soft:max}.  According to this scheme, the best representative model $\hat{z}$ is $\argmax_{\forall z\in\Omega_s}\{\Phi_s(z)\}$. \\
The second scheme is a 2-steps selection model. Firstly a max-stable family is selected:  $\hat{z}$, i.e. $\argmax_{z\in\Omega_m}\{\Phi_m(z)\}$, where $\Omega_{m}=\{\text{Sm, BR, Schl, Geom, Extr}\}$. Then another CNN selects a covariance function in  $\Omega_{\rho}=\{\rho_{exp}, \rho_{gau}, \rho_{chy}, \rho_{wht}\}$. Each of the models  Schlather, Geometric, and Extram-t has corresponding trained CNN to predict the covariance function in  $\Omega_{\rho}$.\\

In the two schemes, the CNNs were trained on the concurrence probability dependence measure, $Con(X(s_i), X(s_j)), (i,j) = 1, \cdots, 30$, defined in \ref{kandall} using the same datasets. Each dataset from the models $\Omega_s$ is $10,000$ times generated, so that  simulate  the 2m air temperature dataset -same numbers and coordinated of the locations and  same number of block-maxima copies-. For each of these datasets, the estimated pairwise concurrence probability for each $ (s,t)=1,\cdots,30$ and the CNNs are trained on these $[30\times 30\times 140k]$ . The layerwise function map  in Equation \ref{layer-func} is
\begin{equation}
g_1:\big(\R^{30\times 30}\otimes\R\big)\times \big(\R^{3\times 3}\otimes\R^{32}\big)\to \big(\R^{28\times 28}\otimes\R^{32}\big).
\end{equation}   
The architectures of the CNNs are the same in the two schemes, except the output layer.  The details of the architectures founded in Table \ref{tab:2} below.

\begin{table}[htb]
\caption{
Architecture of CNNs in the two schemes. The columns from the left to right indicate (respectively) (1) the layer order configured in the CNNs; (2) the type of layers  2D conventional layer, 2D max-pooling, fully connected (dense); (3) Extracted features map to be extracted; (4) the size of the kernel in a convolutional or pooling in max-pooling layers; (5) stride size in convolutional and max -pooling layers; (6) activation function for each trainable layer; (7) the output shape of each layer; (8) the number of trainable parameters $K$ and $W$, respectively corresponding to convolutional and dense layers. In the output layer, the number of parameters is different according to $|\Omega|$. In the first scheme, $|\Omega_s|=14$ leads to parameters equal to $14350$. In the second scheme, the number of parameters in the last layer of the max-stable models CNN  $|\Omega_m|=5$ is $5125 $, while in CNN for the covariances $|\Omega_c|=4$, the number of parameters is also equal to $4100 $.   
  \label{tab:2}}
\small
\begin{tabular}{@{\extracolsep{-1pt}}llcccccc}
 \toprule
 $i$ &   $\mathcal{C}_i$/ $\Psi_i$             & $\ell_i$&$(p_i\times q_i)$&$(\Delta_i\times \Delta^*_i)$ &   $\Phi_i$ & $\R^{n_{i+1}\times m_{i+1}}\otimes\R^{\ell_{i+1}}$& $ K_i / W_i$\\
\midrule
$1$&2D Conv.             & 32         & $3 \times 3$        & $1 \times 1$       & reLU         &      [28, 28, 32] & 320\\
$2$&2D Max-P.           &              & $2 \times 2$        & $1 \times 1$       &                  &      [27, 27, 32]& 0\\
$3$&2D Conv.             & 64         & $3 \times 3$        & $1 \times 1$       &reLU          &      [25, 25, 64] & 18496\\
$4$&2D Max-P.           &              & $2 \times 2$        & $1 \times 1$       &                  &      [24, 24, 64] & 0\\
      &Flatten                &              &                             &                            &                  &      &\\
$5$& Dense               & 256       &                             &                            & geLU         &     [265]&9769225\\
$6$& Dense               & 512       &                             &                            & geLU         &     [512]&136192 \\
$7$& Dense               & 1024     &                             &                            & geLU         &    [1024]& 525312\\
          &Dropout          &              &                             &                            &                   &    &\\
$9$& Dense              & 5           &                             &                            & Softmax     &    [5] & 5125 \\
\midrule
\multicolumn{5}{l}{Total numbers of  trainable parameters}                                             & \multicolumn{1}{l}{}      &&10,454,670      \\
 \bottomrule
\end{tabular}
\end{table}
Kullback-Leibler ($\mathcal{KL}$) divergence is often used for model selection purposes, e.g, see \cite{Cristiano2005}. For that reason, we chose to use $\mathcal{KL}$ defined in Equation \ref{KL}. Adam optimizer with a learning rate tuned at $0.0001$ is chosen for optimization parts. In order to make the training progress more stable,  the $L_2$ regularization  is implemented in each layer with a factor equal to $0.001$. After shuffling in each epoch, $85\%$ from the $140,000$ datasets are devoted to training, and $15\%$  for validation. The training stops when there are no improvements in the loss of the validation dataset. The training of CNNs was executed on the clusters of the Institute Camille Jordan (ICJ – UMR 5208) at Claude Bernard Lyon 1 University.  
In the first scheme, the convolutional neural network denoted by CNN-C is trained to select one of the models in $\Omega_s$.  The metrics of the trained network  CNN-C are  $\mathcal{KL}_c=0.294$ with accuracy equal to $\mathcal{AC}_c=0.931$. Regarding the second scheme, four CNNs are trained, the first one denoted by CNN-M is trained to select one of the models in $\Omega_m$. We get  $\mathcal{KL}_m=0.402$ and accuracy $\mathcal{AC}_m=0.905$. The other convolutional neural networks denoted by CNN-S, CNN-G, and CNN-E are devoted to selecting one of the covariances functions in  $\Omega_{\rho}$. These networks were trained respectively for Schlather, Geomtric and Extremal-t models with metrics: for CNN-S,  $\mathcal{KL}_s=0.087$ $\mathcal{AC}_s=0.999$; CNN-G,    $\mathcal{KL}_g=0.093$ $\mathcal{AC}_g=0.996$; CNN-E, $\mathcal{KL}_e=0.095$ $\mathcal{AC}_e=0.998$.

\section{Results of the model selection methodology}\label{sec:5}
This section is devoted to presenting the results obtained by implementing the CNN approach as an alternative model selection method to select the most representative spatial max-stable process that matches the 2m air temperature dataset. The performance of the trained CNNs in the two schemes is verified by their accuracy in selecting the correct models, compared with the model selection criterion CLIC defined in Equation \ref{clic}. For each fitted model in Section \ref{sec:4},  additional of $1,000$ datasets simulated 2m air temperature were generated. These datasets were not used for the CNNs training and validation. The composite likelihood estimation method was applied to each of the generated datasets. We evaluated the contributions of the composite likelihood for the 14 max-stable spatial models proposed in Section \ref{modeling}. The same as the training process, the selection model by CNNs will be according to the concurrence probability dependence measure.  Tables \ref{tab:3} and \ref{tab:4} present the summary of the accuracies of CLIC and CNNs for the two different schemes. 
\subsection*{Scheme 1}
Table \ref{tab:3} presents the accuracy $\mathcal{AC}$ for the selection of the correct max-stable model in $\Omega_s$ using  CLIC and CNN-C for $1000$ datasets.  Table \ref{tab:6} in the Appendix provides more details for the miss-identify. 

\begin{table}[htb]
\caption{Accuracy of selecting the models in $\Omega_s$ by CNN-C and CLIC, evaluated using $\mathcal{AC}$ on 1,000 datasets generated from the fitted models in  and simulated 2m air temperature dataset. \label{tab:3}}
\centering
\small
\begin{tabular}{@{\extracolsep{-5pt}}lccccccc}
 \toprule
Models  & Smith & BR&Schl-Exp&Schl-Gau & Schl-Chy & Schl-Wht &Geom-Exp \\ 
   \midrule  
CLIC & 0.881 & 0.877 & 0.692 & 0.755 & 0.813 & 0.826 & 0.000 \\ 
  CNN-C & 0.968 & 0.827 & 0.905 & 0.945 & 0.935 & 0.925 & 0.982 \\ 
   \midrule  
Models & Geom-Gau & Geom-Chy&Geom-Wht & Extr-Exp & Extr-Gau &Extr-Cht & Extr-Wht \\ 
    \midrule  
CLIC & 0.904 & 0.424 & 0.360 & 0.850 & 0.808 & 0.899 & 0.807 \\ 
 CNN-C & 0.998 & 0.992 & 0.805 & 0.890 & 0.910 & 0.970 & 0.921 \\ 
 \bottomrule
\end{tabular}
\end{table}
Generally, CNN-C has outperformance in selecting the correct model with non-significant errors than CLIC. This approach can be relied upon in the model selection.  Table \ref{tab:6} in the Appendix indicates that there are significant miss-selections by CLIC for the covariance functions or even models. The CLIC Criterion mis-specifies several models, for example, confused between Schl-Exp (Geom-Exp) and Schl-Wht (Geom-Wht). This could be due to the similarity between  $\rho_{exp}$ and $\rho_{wht}$ when the smooth parameter of the second correlation function is close to 0.5. It is also confused between Geom-Wht with BR since the corresponding stochastic processes in their  spectral representations is  Gaussian  with variogram covariance function. We don't have an explanation for the mis-specification between  Geom-Chy and Geom-Wht.\\

Concerning the 2m air temperature dataset, the predicted probability strength for selecting the models in $\Omega_s$ to $z$ by CNN-C are  $\Phi_s(z\in \text{Schl-Chy})=0.611$, $\Phi_s(z\in \text{Schl-Gau})=0.387$, $\Phi_s(z\in \text{Schl-Wht})=0.001$, and zero for rest models. The best model selected by CNN-C is far from the selected one by CLIC ranked by Table \ref{tab:1} as the eighth. 
 
 \subsection*{Scheme 2}
The accuracies of a model selected by the hierarchical scheme are presented in this part. The same datasets used to train CNN-C were also used in training  CNN-M, CNN-S, CNN-G, and  CNN-E separately.  CNN-M  is dedicated to predicting the dependence strength of matching the models in  $\Omega_m$ to the 2m air temperature dataset. CNN-S, CNN-G, and  CNN-E are trained to select the covariance functions in $\Omega_{\rho}$  corresponding to Schlather, Geometric, and Extremal-t max-stable processes, respectively.  Since CNN-M  and  CNN-C are trained on the same datasets, their performances are more or less equal,  similar to CLIC as shown in Table \ref{tab:4}.  
\begin{table}[htb]
\caption{ Accuracy of selecting the models in $\Omega_m$ by CNN-M and CLIC, evaluated using $\mathcal{AC}$ on 1,000 datasets generated from the fitted models and simulated 2m air temperature dataset. \label{tab:4}}
\centering
\small
\begin{tabular}{@{\extracolsep{4pt}}ll ccccc@{}}
 \toprule
Models &   &Smith & BR &Schlather & Geometric & Extremal-t\\
\midrule
\multirow{2}{*}{Smith}   &CLIC & 0.881 & 0.019 & 0.000 & 0.052 & 0.049 \\ 
&CNN-M & 0.961 & 0.000 & 0.004 & 0.035 & 0.000 \\ 
\midrule
\multirow{2}{*}{BR}&CLIC  & 0.000 & 0.877 & 0.000 & 0.074 & 0.049 \\ 
&CNN-M & 0.000 & 0.780 & 0.005 & 0.003 & 0.212 \\ 
\midrule
\multirow{2}{*}{Schlather} &CLIC& 0.000 & 0.000 & 0.830 & 0.000 & 0.170 \\ 
&CNN-M & 0.006 & 0.001 & 0.898 & 0.089 & 0.006 \\ 
\midrule
\multirow{2}{*}{Geometric} &CLIC  & 0.000 & 0.150 & 0.007 & 0.582 & 0.261 \\ 
&CNN-M & 0.012 & 0.001 & 0.071 & 0.889 & 0.027 \\ 
\midrule
\multirow{2}{*}{Extremal-t}  &CLIC  & 0.000 & 0.000 & 0.005 & 0.000 & 0.995 \\ 
&CNN-M & 0.000 & 0.053 & 0.001 & 0.022 & 0.925 \\ 
 \bottomrule
\end{tabular}
\end{table}

The main advantage of proposing this scheme is the excellent performance of the CNNs corresponding to the selection of the covariance functions. This high performance makes these CNNs a reliable tool with high confidence that can relied upon in the covariance selection problem.  To be fairer in assessing the performance of CLIC in covariance function selection, the accuracies in Table \ref{tab:5} related to each model were executed on the CLICs corresponding to the covariances set $\Omega_{\rho}$ while excluding the rest of the models from this competition.

\begin{table}[htb]
\caption{Accuracies of selecting the covariances in $\Omega_{\rho}$ by CLIC, CNN-S, CNN-G, and CNN-E corresponding to the models  Schlather, Geometric, and  Extremal-t, respectively evaluated using $\mathcal{AC}$ on 1,000 datasets generated from the fitted models simulated 2m air temperature dataset. \label{tab:5}}
\centering
\small
\begin{tabular}{@{\extracolsep{-5pt}}ll|cccc|cccc|cccc}
 \toprule
\multicolumn{2}{@{}c@{}}{} & \multicolumn{4}{@{}c@{}}{Schlather, CNN-S} & \multicolumn{4}{@{}c@{}}{Geometric, CNN-G} & \multicolumn{4}{@{}c@{}}{Extremal-t, CNN-E} \\ 
$\rho$& & Exp & Gau& Chy& Wht&  Exp & Gau& Chy& Wht &  Exp & Gau& Chy& Wht \\ 
\midrule
\multirow{2}{*}{Exp}  &CLIC & 0.828 & 0.000 & 0.009 & 0.163 & 0.098 & 0.000 & 0.000 & 0.902 & 0.851 & 0.000 & 0.005 & 0.144 \\ 
  &CNN & 0.998 & 0.000 & 0.000 & 0.002 & 1.000 & 0.000 & 0.000 & 0.000 & 0.999 & 0.000 & 0.000 & 0.001 \\ 
  \midrule
 \multirow{2}{*}{Gau} &CLIC & 0.000 & 0.910 & 0.032 & 0.058 & 0.000 & 0.918 & 0.010 & 0.072 & 0.000 & 0.817 & 0.108 & 0.075 \\ 
  &CNN & 0.000 & 1.000 & 0.000 & 0.000 & 0.000 & 0.999 & 0.001 & 0.000 & 0.000 & 1.000 & 0.000 & 0.000 \\ 
  \midrule
\multirow{2}{*}{Chy}  &CLIC & 0.000 & 0.000 & 0.977 & 0.023 & 0.000 & 0.001 & 0.431 & 0.568 & 0.017 & 0.000 & 0.903 & 0.080 \\ 
  &CNN & 0.000 & 0.000 & 1.000 & 0.000 & 0.000 & 0.004 & 0.996 & 0.000 & 0.000 & 0.000 & 1.000 & 0.000 \\ 
  \midrule
\multirow{2}{*}{Wht}  &CLIC & 0.000 & 0.000 & 0.013 & 0.987 & 0.005 & 0.000 & 0.039 & 0.956 & 0.025 & 0.000 & 0.165 & 0.810 \\ 
  &CNN & 0.001 & 0.000 & 0.000 & 0.999 & 0.005 & 0.000 & 0.000 & 0.995 & 0.000 & 0.000 & 0.001 & 0.999 \\ 
 \bottomrule
\end{tabular}
\end{table}
As in the first scheme, CLIC faced the same difficulty even when the selection was limited according to each model separately. In this scheme, we will predict the probability strength of matching the covariance in $\Omega_{\rho}$. Identifying the model that matches the dependence structure $z$,  firstly requires identifying the max-stable model in $\Omega_m$ by CNN-M. If the maximum probability strength is one of the models Smith and BR, then the selected model will be one of them. If none of these two models, identifying the covariance function in $\Omega_{\rho}$ will be done by one of CNN-S, CNN-G, and CNN-E according to the model selected previously,  i.e. for 2m air temperature dataset, we have 
\small{
\begin{equation*}
\begin{split}
\max_{z\in\Omega_m}\bigl\{\Phi_m(z)\bigl\}=&\max\Big\{ \Phi_m(z\in \text{Sm}), \Phi_m(z\in \text{BR}),\Phi_m(z\in \text{Schl}),\Phi_m(z\in \text{Geom}),\Phi_m(z\in \text{Extr})\Bigl\}\\
=&\max\{0.000,0.000,1.000,0.000,0.000\}.
\end{split}
\end{equation*}}
That means the max-stable model selected by CNN-M is Schlather with covariance function is $\max_{z \in\Omega_{\rho}}\bigl\{\Phi_{\text{Schl}}(z)\bigl\}$. As  in the first scheme, the ranked models are  $\Phi_{\text{Schl}}(z\in~\text{Shcl-Chy})=0.461$, $\Phi_{\text{Schl}}(z\in~\text{Schl-Gau})=0.300$, and $\Phi_{\text{Schl}}(z\in ~\text{Schl-Wht})=0.239$.

\begin{figure}[htb]
 \centering
\includegraphics[width=1\textwidth]{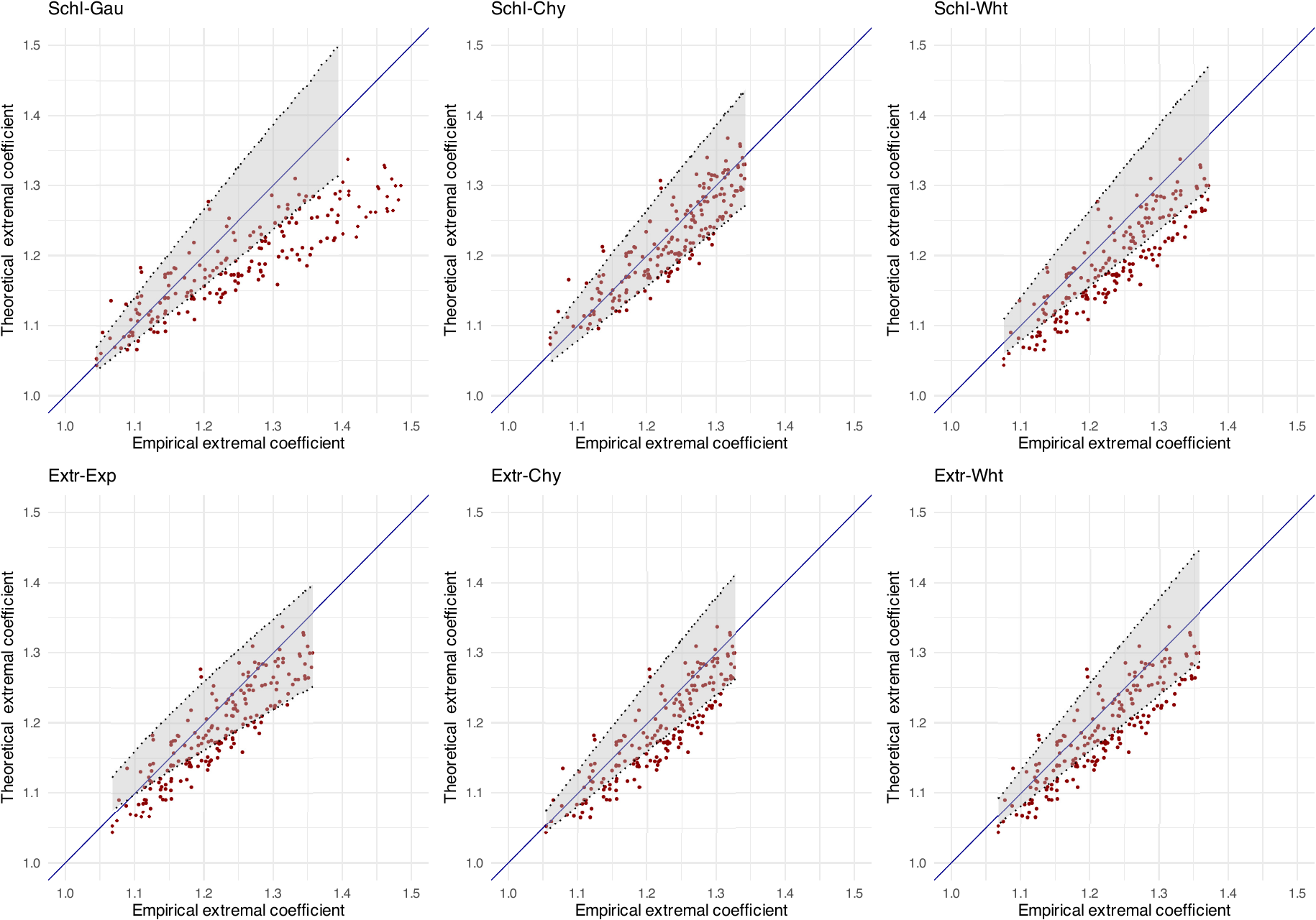}
  \caption{Scatter plots of the theoretical extremal coefficient against the empirical ones. The first row is devoted to models selected by the CNN approach while the second row represents the three best models selected using the CLIC criterion. The grey region is the $95\%$ confidence based on  computed by simulation.   \label{fig:4}}
\end{figure}

The two schemes are selected a model different from those selected by  CLIC. In order to  verify which model selected by CNNs or CLIC best match the 2m air temperature dataset than the ones, we adapt the tools used in \cite{bacro2016flexible} and \cite{oesting2020spatial}. The verification is made of the top three models ranked by CLIC, which are Extr-Exp, Extr-Chy, and Extr-Wht, compared with the models selected by CNNs: Schl-Gau, Schl-Chy, and Schl-Wht. This validation is performed using pairwise extremal coefficients $\theta(s_i,s_j), i,j=\cdots,20$ of the verifying  datasets with, coordinates in white plots in Figure \ref{fig:2}. In Figure \ref{fig:4}, the theoretical extremal coefficients $\theta(||s_i-s_j||,\psi)$ corresponding to each of the top three selected models by CLIC and CNN are scattered against their empirical counterparts $\tilde{\theta}(s_i,s_j)$. The $95\%$ confidence interval of these pairwise extremal coefficients has been constructed from parametric bootstrap extremal coefficients driven from simulated datasets from these six models with $\theta_K, k=1,\cdots,10,000$. \\

First of all, the strength of the extremal coefficients indicates that the dataset demonstrates high spatial dependence, and assesses the nature of the 2m air temperature phenomenon which has a high strength of spatial dependence. Both the Schl-Gau model and Extr-Exp exhibit differences between estimated and empirical extremal coefficients for high and low (resp.) values of the extremal coefficient.  The scatterplot for  Schl-Wht, Ext-Wht and Extr-Chy models, are almost identical and more consistent than the first two models but still present a difference between empirical and theoretical extremal coefficients. Finally,  the most representative model of the dataset seems to be Schl-Chy since most of the pairwise along the dependence strength is inside the uncertainty region.

\section{Discussion and Conclusion}\label{sec:6}
In some cases, the classical model criterion does not exist explicitly or even numerically due to the complexity of the high dimensionality of most environmental problems, especially when dealing with it in a spatial context.  The simulation results demonstrated that the deep learning approach outperforms the most common information criterion CLIC in terms of model selection, especially for the selection of the covariance function.  
\\ \
For the first scheme, CNN-C is trained by $z$ to predict the probability strength of matching the models in $\Omega_s$ in one-step. In the second scheme, the prediction is hierarchical. Firstly, the CNN-M is trained to predict the probability strength of the max-stable models in $\Omega_m$ matching the dataset regardless of the covariance functions. Then, each of CNN-S, CNN-G, and CNN-E are trained to predict the covariances in $\Omega_{\rho}$ matching $z$. The two schemes agreed that the best representative model for the 2m air temperature dataset is Schl-Chy with probability 0.611 and 0.461, respectively. While CLIC identified Extr-Wht as the more representative model among the models in $\Omega_s$. Verifying the best-selected model is implemented by the scatter plots of theoretical extremal coefficients across the empirical ones. Six models are proposed for this purpose: three form the best matching models selected by deep learning and three form the best ranked by CLIC. The confidence interval with $95\%$ of the extremal coefficients $\theta$ corresponding to each selected model is computed by corresponding simulated datasets. The result was in favour of Sch-Chy because most of the plots of pairwise empirical and theoretical extremal coefficients were within the confidence interval.\\
\\ \

\section*{Acknowledgement}
This work was financially supported by the Iraqi Ministry of Higher Education and Scientific Research. We would like to express our gratitude to Université Claude Bernard Lyon 1, ICJ UMR 5208, for their sponsorship and logistical support.



\section*{Funding}

The author Manaf Ahmed is  financially supported by the Iraqi Ministry of Higher Education and Scientific Research.







\bibliographystyle{tfs}
\bibliography{references.bib}

\section*{Appendix}
\appendix

 \begin{sidewaystable}[htb]
\caption{present the details of the accuracies of selecting the models in $\Omega_s$ by CNN-C and CLIC, evaluated using $\mathcal{AC}$ on 1,000 datasets generated from the fitted models in  and simulated 2m air temperature dataset.\label{tab:6}}
\centering
\small
\begin{tabular}{@{\extracolsep{-10pt}}llccccccccccccccc}
   \midrule  
                      &   &    Sm&BR&  Schl-Exp & Schl-Gau & Schl-Chy & Schl-Wht & Geom-Exp& Geom-Gau & Geom-Chy & Geom-Wht& Extr-Exp & Extr-Gau & Extr-Chy & Extr-Wht &  \\ 
  \hline
  \multirow{2}{*}{Smith}  &CLIC & \textbf{0.88} & 0.02 & 0.00 & 0.00 & 0.00 & 0.00  & 0.00 & 0.05 & 0.00 & 0.00 & 0.00 & 0.05 & 0.00 & 0.00  \\ 
                                      &CNN-C & \textbf{0.97} & 0.00 & 0.00 & 0.01 & 0.00 & 0.00  & 0.00 & 0.00 & 0.00 & 0.00& 0.00 & 0.02 & 0.00 & 0.00  \\ 
   \midrule  
   \multirow{2}{*}{BR}  & CLIC & 0.00 & \textbf{0.88} & 0.00 & 0.00 & 0.00 & 0.00 & 0.00 & 0.00 & 0.00 & 0.07& 0.04 & 0.00 & 0.00 & 0.01  \\ 
                                   & CNN-C & 0.00 &\textbf {0.83} & 0.00 & 0.00 & 0.00 & 0.01 & 0.00 & 0.00 & 0.00 & 0.17& 0.00 & 0.00 & 0.00 & 0.00  \\ 
   \midrule  
    \multirow{2}{*}{Schl-Exp}  &CLIC& 0.00 & 0.00 & \textbf{0.69 }& 0.00 & 0.01 & 0.27 & 0.00 & 0.00 & 0.00 & 0.00& 0.00 & 0.00 & 0.00 & 0.04  \\ 
                                             & CNN-C & 0.00 & 0.00 &\textbf{ 0.91} & 0.00 & 0.00 & 0.00 & 0.00 & 0.00 & 0.00 & 0.00& 0.09 & 0.00 & 0.00 & 0.00  \\ 
   \midrule  
    \multirow{2}{*}{Schl-Gau}  &CLIC & 0.00 & 0.00 & 0.00 & \textbf{0.76 }& 0.03 & 0.04 & 0.00 & 0.00 & 0.00 & 0.00& 0.00 & 0.14 & 0.01 & 0.01  \\ 
                                              &CNN-C & 0.01 & 0.00 & 0.00 &\textbf{0.94 }& 0.00 & 0.00 & 0.00 & 0.00 & 0.00 & 0.00 & 0.00 & 0.04 & 0.00 & 0.00 \\ 
   \midrule  
        \multirow{2}{*}{Schl-Chy}  &CLIC & 0.00 & 0.00 & 0.00 & 0.00 & \textbf{0.81} & 0.02 & 0.00 & 0.00 & 0.00 & 0.00 & 0.00 & 0.00 & 0.16 & 0.00 \\ 
                                             & CNN-C & 0.00 & 0.00 & 0.00 & 0.00 & \textbf{0.94} & 0.00 & 0.00 & 0.00 & 0.02 & 0.00  & 0.00 & 0.00 & 0.05 & 0.00 \\ 
   \midrule  
    \multirow{2}{*}{Schl-Wht}  &CLIC& 0.00 & 0.00 & 0.00 & 0.00 & 0.01 & \textbf{0.83} & 0.00 & 0.00 & 0.00 & 0.00& 0.00 & 0.00 & 0.01 & 0.16  \\ 
                                              &CNN-C & 0.00 & 0.00 & 0.00 & 0.00 & 0.00 & \textbf{0.93}& 0.00 & 0.00 & 0.00 & 0.00& 0.00 & 0.00 & 0.00 & 0.07  \\ 
   \midrule  
    \multirow{2}{*}{Geom-Exp}  &CLIC& 0.00 & 0.00 & 0.00 & 0.00 & 0.03 & 0.00 & \textbf{0.00 }& 0.00 & 0.00 &0.96& 0.00 & 0.00 & 0.01 & 0.00  \\ 
                                             & CNN-C & 0.00 & 0.00 & 0.00 & 0.00 & 0.00 & 0.00  & \textbf{0.98} & 0.00 & 0.00 & 0.00 & 0.02 & 0.00 & 0.00 & 0.00\\ 
   \midrule  
    \multirow{2}{*}{Geom-Gau}  &CLIC & 0.00 & 0.00 & 0.00 & 0.00 & 0.00 & 0.00 & 0.00 & \textbf{0.90} & 0.01 & 0.07& 0.01 & 0.00 & 0.00 & 0.00  \\ 
                                             &CNN-C & 0.00 & 0.00 & 0.00 & 0.00 & 0.00 & 0.00 & 0.00 & \textbf{1.00} & 0.00 & 0.00 & 0.00 & 0.00 & 0.00 & 0.00 \\ 
   \midrule  
    \multirow{2}{*}{Geom-Chy}  &CLIC & 0.00 & 0.00 & 0.00 & 0.00 & 0.00 & 0.00 & 0.00 & 0.00 & \textbf{0.42 }& 0.56 & 0.02 & 0.00 & 0.00 & 0.00 \\ 
                                             &CNN-C & 0.00 & 0.00 & 0.00 & 0.00 & 0.01 & 0.00 & 0.00  & 0.00 & \textbf{0.99 }& 0.00 & 0.00 & 0.00 & 0.00 & 0.00  \\ 
   \midrule  
    \multirow{2}{*}{Geom-Wht}  &CLIC& 0.00 & 0.60 & 0.00 & 0.00 & 0.00 & 0.00 & 0.00 & 0.00 & 0.00 & \textbf{0.36} & 0.03 & 0.00 & 0.00 & 0.01 \\ 
                                             &CNN-C & 0.00 & 0.18 & 0.00 & 0.00 & 0.00 & 0.00 & 0.00 & 0.00 & 0.00 &\textbf{ 0.81} & 0.00 & 0.00 & 0.00 & 0.01 \\

   \midrule  
   \multirow{2}{*}{Extr-Exp}  & CLIC & 0.00 & 0.00 & 0.00 & 0.00 & 0.00 & 0.00 & 0.00 & 0.00 & 0.00 & 0.00& \textbf{0.85} & 0.00 & 0.01 & 0.14  \\ 
                                               &CNN-C & 0.00 & 0.00 & 0.09 & 0.00 & 0.00 & 0.00 & 0.02 & 0.00 & 0.00 & 0.00 & \textbf{0.89} & 0.00 & 0.00 & 0.00 \\ 
   \midrule  
    \multirow{2}{*}{Extr-Gau}  &CLIC & 0.00 & 0.00 & 0.00 & 0.01 & 0.00 & 0.00 & 0.00 & 0.00 & 0.00 & 0.00 & 0.00 &\textbf{ 0.81} & 0.11 & 0.07 \\ 
                                                & CNN-C & 0.04 & 0.00 & 0.00 & 0.05 & 0.00 & 0.00 & 0.00 & 0.00 & 0.00 & 0.00 & 0.00 &\textbf{ 0.91} & 0.00 & 0.00 \\ 
   \midrule  
   \multirow{2}{*}{Extr-Chy}  &CLIC & 0.00 & 0.00 & 0.00 & 0.00 & 0.00 & 0.00 & 0.00 & 0.00 & 0.00 & 0.00 & 0.02 & 0.00 &\textbf{0.90 }& 0.08 \\ 
                                              & CNN-C & 0.00 & 0.00 & 0.00 & 0.00 & 0.03 & 0.00 & 0.00 & 0.00 & 0.00 & 0.00 & 0.00 & 0.00 & \textbf{0.97} & 0.00 \\ 
   \midrule  
    \multirow{2}{*}{Extr-Wht}  &CLIC & 0.00 & 0.00 & 0.00 & 0.00 & 0.00 & 0.00 & 0.00 & 0.00 & 0.00 & 0.00& 0.03 & 0.00 & 0.17 & \textbf{0.81}  \\ 
                                              & CNN-C & 0.00 & 0.00 & 0.00 & 0.00 & 0.00 & 0.05 & 0.00 & 0.00 & 0.00 & 0.03 & 0.00 & 0.00 & 0.00 & \textbf{0.92}\\ 
   
   \midrule  
\end{tabular}
\end{sidewaystable}
\end{document}